\documentclass[amsmath,amssymb,aps,prl,twocolumn]{revtex4-1}

\usepackage{amssymb}
\usepackage{graphicx}
\usepackage{dcolumn}
\usepackage{bm}
\usepackage{amsmath}
\usepackage{mathrsfs}
\usepackage{upgreek}

\usepackage{textcomp}
\usepackage[mathlines]{lineno}

\usepackage[unicode=true,bookmarks=true,bookmarksnumbered=false,bookmarksopen=false,breaklinks=false,pdfborder={0 0 1},backref=false,colorlinks=true]{hyperref}
\hypersetup{linkcolor=magenta,urlcolor=blue,citecolor=blue,pdfstartview={FitH},unicode=true}

\begin{document}
\preprint{APS/123-QED}

\title{Homogeneous fermionic Hubbard gases in a flat-top optical lattice}

\author{Yu-Xuan Wang$^{1,2}$}
\thanks{These authors contributed equally to this work.}
\author{Hou-Ji Shao$^{1,2}$}
\thanks{These authors contributed equally to this work.}
\author{Yan-Song Zhu$^{1,2}$}
\author{De-Zhi Zhu$^{1,2}$}
\author{Hao-Nan Sun$^{1,2}$}
\author{Si-Yuan Chen$^{1,2}$}
\author{Xing-Can Yao$^{1,2,3}$}
\thanks {yaoxing@ustc.edu.cn}
\author{Yu-Ao Chen$^{1,2,3}$}
\thanks{yuaochen@ustc.edu.cn}
\author{Jian-Wei Pan$^{1,2,3}$}
\thanks{pan@ustc.edu.cn}

\affiliation{$1$Hefei National Research Center for Physical Sciences at the Microscale and School of Physical Sciences, University of Science and Technology of China, Hefei 230026, China}
\affiliation{$2$Shanghai Research Center for Quantum Science and CAS Center for Excellence in Quantum Information and Quantum Physics, University of Science and Technology of China, Shanghai 201315, China}
\affiliation{$3$Hefei National Laboratory, University of Science and Technology of China, Hefei 230088, China}


\begin{abstract}
	Fermionic atoms in a large-scale, homogeneous optical lattice provide an ideal quantum simulator for investigating the fermionic Hubbard model, yet achieving this remains challenging. Here, by developing a hybrid potential that integrates a flat-top optical lattice with an optical box trap, we successfully realize the creation of three-dimensional, homogeneous fermionic Hubbard gases across approximately $8\times10^5$ lattice sites. This homogeneous system enables us to capture a well-defined energy band occupation that aligns perfectly with the theoretical calculations for a zero-temperature, ideal fermionic Hubbard model. Furthermore, by employing novel radio-frequency spectroscopy, we precisely measure the doublon fraction $D$ as a function of interaction strength $U$ and temperature $T$, respectively. The crossover from metal to Mott insulator is detected, where $D$ smoothly decreases with increasing $U$. More importantly, we observe a non-monotonic temperature dependence in $D$, revealing the Pomeranchuk effect and the development of extended antiferromagnetic correlations.  
\end{abstract}

\maketitle

Ultracold fermionic atoms in optical lattices\cite{bloch2005ultracold, esslinger2010fermi}, owing to their unparalleled purity and excellent controllability, hold great promise for exploring the fermionic Hubbard (FH) physics\cite{arovas2022hubbard}. To this end, it is imperative to realize a homogeneous FH system, where the nearest-neighbor hopping $t$, the on-site interaction $U$ between atoms with opposite spins, and the local chemical potential $\mu$ remain nearly constant across the system. This is particularly crucial at low temperatures, as changes in these Hubbard parameters can profoundly influence the physical properties of the system. Unfortunately, in conventional optical lattices created by the Gaussian laser beams, all the Hubbard parameters exhibit significant spatial variations. This not only hinders the observation of exotic quantum phases but also makes experimental results difficult to serve as benchmarks for theoretical and numerical studies\cite{qin2022hubbard}.

To address the issue of system inhomogeneity, various compensation methods have been developed to mitigate the underlying harmonic confinement in the Gaussian optical lattices\cite{simon2011quantum,schneider2012fermionic, hart2015observation,mazurenko2017cold}. While these techniques alleviate the spatial variation of $\mu$, they fail to tackle the spatial dependence of $t$ and $U$. Nevertheless, in two-dimensional (2D) optical lattices, the advent of quantum gas microscopes\cite{gross2017quantum, gross2021quantum} has enabled the isolation and detection of the central region of the system, leading to the realization of homogeneous, albeit small, systems\cite{mazurenko2017cold, brown2019bad, hartke2023direct}. Unfortunately, in three-dimensional (3D) optical lattices, the rapid divergence of lasers with beam waists comparable to the lattice spacing poses a great challenge in isolating and investigating a small, homogeneous system. Consequently, the measurement of any physical observables yields averaged results across different parameter settings, often obscuring important yet subtle physical phenomena, such as the elusive Pomeranchuk effect in SU(2) FH systems\cite{pomeranchuk1950He, richardson1997pomeranchuk}. Even for more noticeable phenomena, like the evolution of the double occupancy (doublon) fraction with interaction strength and temperature\cite{schneider2008metallic, jordens2008mott, de2008trapping, scarola2009discerning, gorelik2010neel, greif2016site, drewes2016thermodynamics}, intricate comparisons between experiment and theory are required to establish a proper understanding, i.e., employing the Local Density Approximation (LDA) to translate localized numerical results to those applicable under inhomogeneous conditions and compare with experimental data. 

\begin{figure}[tbp]
	\includegraphics[width=1\columnwidth]{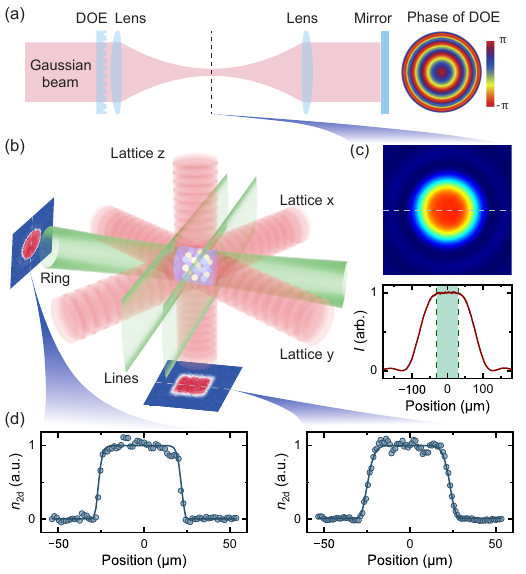}
	\caption{\label{fig1} Schematic experimental setup. (a) Sketch of the 1D flat-top optical lattice. A Gaussian laser beam passes through a custom-designed diffractive optical element (DOE) and an achromatic lens ($f=200$~mm) to produce a laser beam with a uniform-intensity distribution and a flat wavefront at the position of the atoms. The beam is then collimated by another achromatic lens and retro-reflected by a mirror to form the optical lattice. The right panel shows the phase pattern of the DOE. (b) Sketch of the hybrid potential, consisting of three mutually orthogonal 1D flat-top optical lattices and a cylindrical optical box trap comprised of a ring and two line beams. The white and blue balls represent atoms with opposite spins, respectively, accompanied by two images showing their column-integrated \textit{in situ} density distributions from two perpendicular directions. (c) Upper panel: the 2D intensity profile $I$ of the flat-top laser. Lower panel: the 1D intensity cut along the white dashed line in $I$, with a green shaded band indicating the central uniform region, which has a diameter of $60~\upmu$m. (d) Cuts through the column-integrated density profiles. Solid lines are obtained by fitting the experimental dots to super-Gaussian curves.}
\end{figure}

In this Letter, we report on the realization and investigation of a large, homogeneous 3D FH system within a flat-top optical lattice (FOL). This is achieved through a hybrid potential that combines a red-detuned 3D FOL\cite{shao2024antiferromagnetic} with a blue-detuned optical box trap\cite{navon2021quantum, li2022second, li2024observation}, as shown in Fig.~\ref{fig1}. More specifically, the FOL, characterized by a uniform intensity profile across its central region, with a diameter of approximately 60 $\upmu$m (e.g., see Fig.~\ref{fig1}(c)), is generated through the modulation of the spatial phases\cite{soifer2002methods} of Gaussian lattice beams using diffractive optical elements (DOE), as illustrated in Fig.~\ref{fig1}(a). The phase profile of the DOE, shown in the right panel of Fig.~\ref{fig1}(a), can be designed using an iterative Fourier transform algorithm\cite{2004Iterative} and etched onto a fused silica window via inductively coupled plasma technology. Integration with a cylindrical box trap, this approach not only isolates this highly uniform central region but also prevents atoms from extending into the less homogeneous outer regions of the FOL. As a benchmark, we measure quasi-momentum distributions of non-interacting, homogeneous FH gases as a function of lattice filling $n$, using band mapping technique. We clearly observe a well-defined Fermi surface and quantify its expansion with increasing $n$, which agrees well with theoretical predictions for a zero-temperature, homogeneous FH system. Moreover, by developing a radio-frequency (RF) spectroscopy, we precisely measure the doublon fraction ($D$) of a half-filled FH system, varying interaction strength and temperature. With increasing interaction strength, we observe a smooth decrease in $D$, manifesting the crossover from metallic to Mott insulating regimes\cite{jordens2008mott, boll2016spin, kim2020spin}. Remarkably, an anomalous increase in $D$ is unequivocally revealed as the entropy per particle $s$ decreases in the $0.4k_\text{B}\lesssim s\lesssim 0.9k_\text{B}$ regime, which is a hallmark of Pomeranchuk effect\cite{werner2005interaction, de2011thermodynamics, sciolla2013competition}. In the vicinity of the lowest entropy, i.e., $s\lesssim 0.4k_\text{B}$, we observe a decrease in $D$ as entropy further reduces, signaling the approaching of an antiferromagnetic phase transition\cite{shao2024antiferromagnetic}. These findings highlight the advantages of the large, homogeneous FH gases as a valuable platform for exploring FH physics.

Our experiment is based on a balanced spin mixture of fermionic $^6$Li atoms\cite{yao2016observation, liu2022temperature} in the hyperfine levels $|1\rangle \equiv |F=\frac{1}{2}, m_F=\frac{1}{2}\rangle$ and $|3\rangle \equiv |F=\frac{3}{2}, m_F=-\frac{3}{2}\rangle$, confined in a 3D FOL within a cylindrical box trap (see Fig.~\ref{fig1}(b)). The 3D FOL is formed by intersecting three mutually orthogonal 1D optical lattices, each created by two counter-propagating flat-top laser beams at 1064~nm. The cylindrical box trap, generated by a ring (diameter of 49.5~$\upmu$m) and two line 532~nm laser beams (separation of 47.7~$\upmu$m), defines the system's boundary and isolates approximately $8\times10^5$ lattice sites. 

We begin by preparing a near-degenerate homogeneous Fermi gas in the box trap. To achieve deep quantum degeneracy, we ramp down the potential depth with an optimized evaporative cooling curve for 3.5~s at 390~G, and then linearly increase the trap depth to $13.5~\upmu$K within 100~ms to isolate the gas further. Thermometry is performed by ramping the magnetic field from 390~G to 568.14~G, where the $s$-wave scattering length between states $|1\rangle$ and $|3\rangle$ vanishes. This is followed by measuring the system's momentum distribution using time-of-flight imaging and fitting the data to a Fermi-Dirac distribution. We obtain a quite low temperature of $T/T_\text{F} = 0.042(1)$ at a density of approximately $6.64\times10^{12}/\text{cm}^3$ (corresponding to half filling, i.e., $n=1$, in the homogeneous optical lattice), where $T_\text{F}$ is the Fermi temperature. Taking into account the actual shape of the box trap, characterized by edge thicknesses with $1/e^2$ radii of 2.35$~\upmu$m (ring beam) and 3.6$~\upmu$m (line beams), respectively, the initial entropy per particle, $s_0$, of the homogeneous Fermi gas is calculated to be $s_0= 0.234(4)k_\text{B}$, where $k_\text{B}$ is the Boltzmann constant.

Subsequently, the homogeneous Fermi gas is loaded into the FOL by increasing the lattice depth to $6.25 E_\text{r}$ through a three-stage, 18~ms process, achieving a nearest-neighbor hopping $t/h = 1.40(1)$~kHz: first to $0.5 E_\text{r}$ in 6~ms, then to $2.4 E_\text{r}$ in another 6~ms, and finally to $6.25 E_\text{r}$ in the last 6~ms, where $E_\text{r}/h=29.30$~kHz is the recoil energy. Simultaneously, the magnetic field is ramped to a target value within the first 12~ms and held for the remaining 6~ms to adiabatically reach the desired onsite interaction $U$, which can be any value $0\leqslant U/h\leqslant 24.67(8)$~kHz. Figure~\ref{fig1} presents the exemplary \textit{in situ} images of the FH gases at half filling, where uniform density distributions across a diameter of approximately $40~\upmu$m are clearly visible in the 1D density cuts.

\begin{figure}[tbp]
	\includegraphics[width=1\columnwidth]{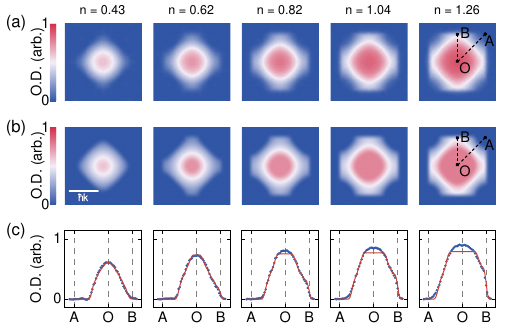}
	\caption{\label{fig2} Integrated quasi-momentum distributions of non-interacting homogeneous FH gases. (a) The measured 2D quasi-momentum distributions for $n = 0.43$, $0.62$, $0.82$, $1.04$, and $1.26$, respectively, with each image corresponding to an average of approximately 80 raw images. (b) Theoretical integrated quasi-momentum distributions, calculated by applying zero-temperature Fermi-Dirac distributions to Bloch eigenstates, and convolving them with the initial atomic density profiles. (c) Quasi-momentum cuts. The blue dots are measured cuts along $q_x = q_y$ (O to A) and $q_y = 0$ (O to B), while the red solid lines correspond to theoretical predictions. }
\end{figure}

We first investigate the quasi-momentum distribution for the non-interacting case ($U=0$), since solid theoretical predictions exist and can serve as a benchmark for our homogeneous FH system. Employing band mapping, specifically, adiabatically ramping down the lattice potential in 100~$\upmu$s and allowing for a 4~ms ballistic expansion after switching off the box trap, we probe the 2D optical density profile of the atoms, which corresponds to the integrated quasi-momentum distribution along the imaging axis. We focus on the characterization of the Fermi surface, which distinguishes occupied from unoccupied states in reciprocal space, and is essential for elucidating energy band occupation, the lattice's periodicity and symmetry, and strong correlation effects within the FH system. As shown in Fig.~\ref{fig2}(a), sharp edges in quasi-momentum distribution are distinctly visible, a defining feature of the Fermi surface. Moreover, the measured quasi-momentum distributions closely resemble the theoretical calculations for the FH system at zero temperature, as shown in Fig.~\ref{fig2}(b), over a wide range of $0.43\leq n\leq 1.26$. These observations starkly contrast with those in the inhomogeneous case, where the spatial variation in lattice potential not only broadens and blurs the Fermi surface but also complicates the accurate determination of the energy band occupation\cite{rigol2004quantum, kohl2005fermionic}. We mention that, due to the finite duration of the time-of-flight, the influence of the initial cloud size on the measured quasi-momentum distributions cannot be neglected, which slightly smooths the sharp edges and increases the central weight of the quasi-momentum distribution.

Quantitative comparisons between experiment and theory are presented in Fig.~\ref{fig2}(c). For low lattice fillings, specifically at $n = 0.43$ and 0.62, excellent agreement between experimental data and theoretical curves is achieved, demonstrating the homogeneity and low temperature of the experimental FH system. For higher lattice fillings, particularly at $n=1.26$, although good accordance is obtained, a noticeable discrepancy between experiment and theory emerges near the center and edges of the Brillouin zone. This is because, a large number of atoms occupy states close to the Brillouin zone edges, where the energy gaps between $s$ and $p$ bands are minimal and progressively decrease with the reduction of lattice potential during the band mapping process\cite{mckay2009lattice}. Such conditions facilitate non-adiabatic inter-band excitations, distorting the sharp edges in the quasi-momentum distribution. These edge imperfections are further manifested in the column density at the center as a result of integration along the imaging axis, leading to the deviation from an ideal flat-top distribution.

We then address the main focus of this work: an interacting homogeneous FH system, by probing the fraction of doublon, i.e., two atoms with opposite spins occupy the same lattice site, across various Hubbard parameters. The doublon fraction $D = 2\langle n_\uparrow n_\downarrow\rangle / (\langle n_\uparrow \rangle + \langle n_\downarrow \rangle)$ plays a crucial role in understanding FH physics, including band and Mott insulating phenomena\cite{schneider2008metallic, jordens2008mott}, thermodynamic properties\cite{jordens2010quantitative, cocchi2016equation}, local magnetic correlations\cite{greif2011probing, boll2016spin}, and the pairing of fermions\cite{hartke2023direct}. We develop a novel method, i.e., radio-frequency spectroscopy\cite{campbell2006imaging, li2024observation, 2024Thermography} in the linear response regime, to determine $D$ with higher accuracy and resolution, compared to existing techniques\cite{taie20126,schneider2008metallic,cocchi2016equation,jordens2010quantitative}, especially when $D$ is small. In the measurement stage, the lattice potential depth is rapidly increased to 20$E_\text{r}$ in 20~$\upmu$s to freeze the atoms, allowing the system to be treated as isolated lattice sites. Then, a small portion of atoms in level $|1\rangle$ are transferred to an initially unoccupied hyperfine level $|2\rangle \equiv |F=\frac{1}{2}, m_F=-\frac{1}{2}\rangle$ using a Gaussian-shaped microwave pulse. The total and $1/\sqrt e$ half-width pulse durations are 800 and 170~$\upmu$s, respectively, leading to a Fourier-limited  energy resolution of approximately $h\times 1.4$~kHz. Finally, we measure the number of atoms $N_2$ in level $|2\rangle$ versus the RF frequency $\nu$ using absorption imaging. 

\begin{figure}[tbp]
	\includegraphics[width=8.9cm]{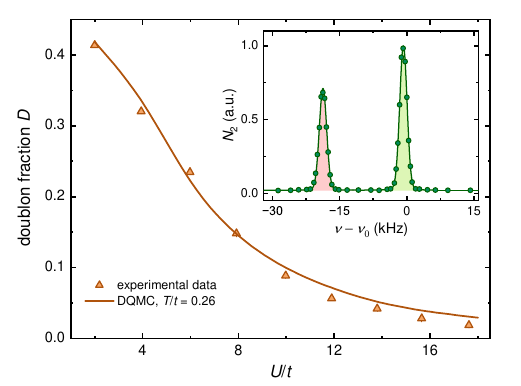}
	\caption{
		Doublon fraction $D$ as a function of $U$ at half filling, and with an entropy per particle $s \simeq 0.3\text{k}_B$. Error bars on $D$ indicate one standard error. The solid orange line represents the value of $D$ obtained from DQMC simulations for a system size $L=6$ at half filling and $T/t = 0.26$. Inset: radio-frequency spectroscopy for determining $D$. The normalized number of atoms in level $|2\rangle$ is plotted against the RF frequency offset $\nu-\nu_0$, where $\nu_0=75754.407$~kHz corresponds to resonant RF frequency of single occupancy. Each data point represents the average of $6$ measurements. The solid green line depicts a double-peak fit where each peak is represented by a linear superposition of Gaussian and Lorentzian functions.}
	\label{fig3}
\end{figure}

As an illustration, we present the measured spectrum at $U=2.00(5)t$ and half filling in the inset of Fig.~\ref{fig3}, along with a fit to a double-peak function, where each peak is represented by a linear combination of Gaussian and Lorentz functions. Due to the strong on-site interaction at 20$E_\text{r}$, two distinct peaks are observed, with the higher and lower peaks corresponding to the spectral responses of single and double occupancies, respectively. The energy separation between these peaks approximately satisfies the relation $\nu_\text{double} - \nu_\text{single} \simeq U_{23} - U_{13}$, where $U_{23}$ and $U_{13}$ denote the on-site interactions of $|2\rangle$-$|3\rangle$ and $|1\rangle$-$|3\rangle$ doublons, respectively. According to the sum rule $\int N_2(\nu)\mathrm{d}\nu=S_1+S_2\propto N_1$, where $N_1$ is the total number of atoms in level $|1\rangle$, the doublon fraction $D=0.414\pm0.003$ is determined by the ratio $S_2/(S_1+S_2)$. Here, $S_1$ (green shaded region) and $S_2$ (red shaded region) represent the spectral areas corresponding to single and double occupancies, respectively.

\begin{figure*}[htbp]
	\includegraphics{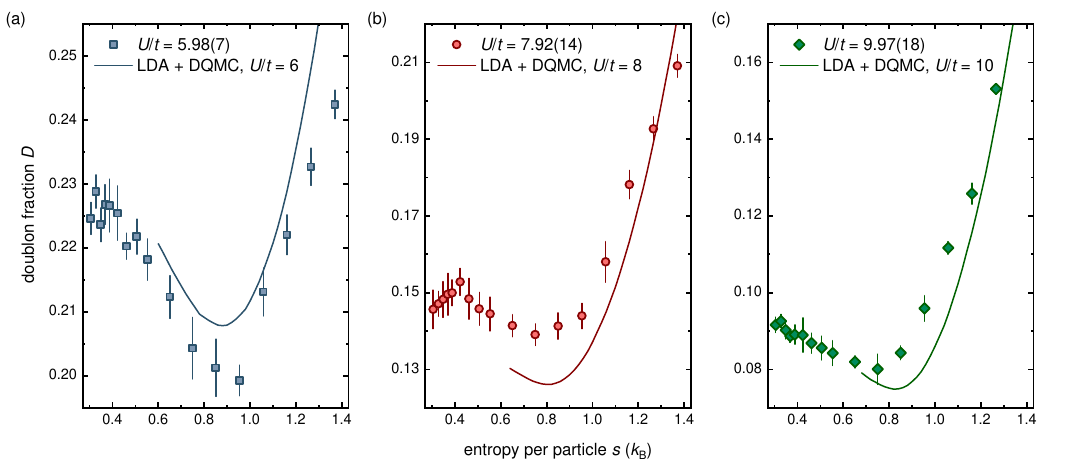}
	\caption{\label{fig4}
		The doublon fraction $D$ as a function of entropy per particle $s$ at half filling for $U = 5.98(7)$, $7.92(14)$, and $9.97(18)t$, respectively. Error bars represent one standard error. The solid lines represent LDA results that account for the boundary effects of the box trap. These results are obtained using DQMC data for a linear lattice size of $L=6$, which includes Gaussian-distributed disorder in the chemical potential with a standard deviation of $\sigma_\mu = 0.5t$.}
\end{figure*}

Using this method, we measure the doublon fraction $D$, as a function of interaction strength $U$ at half filling with an entropy per particle $s \simeq 0.3k_\text{B}$, determined from the round-trip entropy thermometry. Note that, The entropy per particle is calculated as $s = s_i \times \sqrt{s_f/s_i}$, where $s_i$ and $s_f$ denote the measured entropy per particle before and after the round-trip process, under the assumption that the entropy increase during the ramping up or down of lattice potential is the same. The results are displayed in Fig.~\ref{fig3}. As anticipated, we observe a rapid yet smooth decrease in $D$ with increasing $U$, which closely aligns with Determinant Quantum Monte Carlo (DQMC) results at a temperature of $0.26t$ and half filling. For small values of $U$, the doublon fraction is high, indicating that the system is in a metallic state with significant charge fluctuations\cite{drewes2016thermodynamics}. Conversely, for large values of $U$, $D$ decreases to approximately 0.5\% and approaches zero as $U$ increases further, signifying the localization of the atoms and the formation of a Mott insulator\cite{jordens2008mott,de2011thermodynamics}. These findings demonstrate that, in contrast to inhomogeneous systems--where dependencies of $D$ on $n$ and $U$ are significantly affected by the detailed experimental parameters\cite{schneider2008metallic, de2008trapping}--measurements in a homogeneous FH system can uncover physical phenomena more directly.

We now explore the temperature dependence of $D$ for various interaction strengths at half filling. To prepare FH system at higher temperatures, we heat the coldest homogeneous Fermi gas for varying periods by exposing the atoms to an extremely weak and far-detuned laser, followed by a standard lattice loading procedure. The experimental results for $U = 5.98(7)$, $7.92(14)$, and $9.97(18)t$ are shown in Fig.~\ref{fig4}. It is important to note that our experimental system is susceptible to imperfections, including deviations from half filling at the boundaries and residual disorder in the lattice potential. Although these imperfections are small, they exert a non negligible influence on the doublon fraction\cite{shao2024antiferromagnetic}. To account for these effects, we first perform DQMC simulations with Gaussian distributed disorder in the chemical potential, characterized by a standard deviation of $\sigma_\mu= 0.5t$. These simulations are conducted at $T>0.5t$, where the sign problem is less severe. Our findings indicate that residual disorder slightly increases the doublon fraction $D$ across the entire temperature range but does not alter its temperature dependence. Utilizing these DQMC data and precise knowledge of the box trap potential, we further apply the LDA method to calculate the doublon fraction, represented by the solid lines in Fig.~\ref{fig4}, which show good agreement with the experimental results. 

Strikingly, at $U = 5.98(7)t$, we identify three distinct phases as the initial entropy per particle $s$ reduced. In the high entropy regime, as anticipated, $D$ decreases with $s$ due to the suppression of thermal-induced charge fluctuations. At $s\simeq 0.9k_{\text{B}}$, $D$ reaches a local minimum. Remarkably, both experimental and numerical results reveal an anomalous increase in $D$ when $s$ further decreases. This intriguing phenomenon, reminiscent of the Pomeranchuk effect\cite{pomeranchuk1950He} observed in Helium-3, where a solid counter-intuitively liquefies upon cooling, arises from the presence of excess spin entropy in the localized states. As the temperature lowers, spin entropy reduces more rapidly than the total entropy, leading to an increase in charge fluctuations as a compensatory mechanism. We mention that Pomeranchuk cooling has been observed in SU($N$) FH systems\cite{taie20126, hofrichter2016direct}, due to their large spin degrees of freedom and high spin entropy. However, for SU(2) FHM, this effect has not been observed before, despite theoretical predictions made nearly two decades ago\cite{werner2005interaction}, largely due to the absence of a large and homogeneous system.

At $s\simeq 0.4k_{\text{B}}$, $D$ attains a saturated value, indicating the saturation of local spin correlations and the emergence of extended antiferromagnetic correlations. A similar non-monotonic dependence of $D$ on temperature is also obtained for $U = 7.92(14)t$, although with a less pronounced Pomeranchuk effect in the $0.42k_\text{B} \lesssim s \lesssim 0.75k_\text{B}$ regime compared to the case of $U = 5.98(7)t$. Additionally, since the N\'eel temperature is higher at\cite{kozik2013neel, qin2022hubbard} $U \simeq 8t$, the system at its lowest temperature is closer to the antiferromagnetic phase transition\cite{shao2024antiferromagnetic}. This proximity leads to an observable decrease in the measured doublon fraction near the lowest $s$. At a stronger interaction strength of $U = 9.97(18)t$, $D$ gradually approaches a minimal value of approximately 0.08 at $s \simeq 0.75k_{\text{B}}$. Subsequently, $D$ slightly increases as $s$ continues to decreases, where doublons induced by thermal fluctuations are significantly reduced, leaving quantum fluctuations as the primary mechanism for doublon creation.

In summary, we successfully create a homogeneous, large-scale, and low-temperature fermionic Hubbard system by confining fermionic $^6$Li atoms in a novel hybrid optical potential that combines a flat-top lattice with a box trap. This breakthrough enables us to explore the FH physics in previously inaccessible parameter regimes and to reveal intriguing physics from the experimental data directly, leading to the observation of a nearly perfect Fermi surface and the unveiling of a fascinating interplay between spin and charge fluctuations. In the near future, we aim to observe and investigate the fermion interference within the realized homogeneous lattice, which is free from inter-band excitations suffered in band mapping, thus enabling the acquisition of more accurate quasi-momentum distributions. Moreover, with the ability to precisely measure $D$, our setup is particularly suited for performing lattice modulation spectroscopy\cite{greif2011probing, sensarma2009modulation}, which allows for the identification of antiferromagnetic phase transition\cite{tokuno2012spin, sensarma2009modulation}, the probing of quasiparticle excitations\cite{timusk1999pseudogap}, and the exploration of possible $d$-wave superfluidity\cite{xiang2022d}.

We thank Y. Deng and Y.-Y. He for discussions. This work is supported by the Innovation Program for Quantum Science and Technology (Grant No. 2021ZD0301900), NSFC of China (Grant No. 11874340), the Chinese Academy of Sciences (CAS), the Anhui Initiative in Quantum Information Technologies, the Shanghai Municipal Science and Technology Major Project (Grant No.2019SHZDZX01), and the New Cornerstone Science Foundation.

\end{document}